\documentclass[aip,reprint]{revtex4-1}
\usepackage{xcolor}
\usepackage{graphicx}
\usepackage{float}
\usepackage{mhchem}

\draft 

\begin{document}

\title{A Semiconductor Topological Photonic Ring Resonator} 

\author{M. Jalali Mehrabad}
\email[]{mjalalimehrabad1@sheffield.ac.uk}
\affiliation{Department of Physics and Astronomy, University of Sheffield, Sheffield S3 7RH, UK}

\author{A.P. Foster}
\email[]{andrew.foster@sheffield.ac.uk}
\affiliation{Department of Physics and Astronomy, University of Sheffield, Sheffield S3 7RH, UK}

\author{R. Dost}
\affiliation{Department of Physics and Astronomy, University of Sheffield, Sheffield S3 7RH, UK}

\author{E. Clarke}
\affiliation{EPSRC National Epitaxy Facility, University of Sheffield, Sheffield S1 4DE, UK}

\author{P.K. Patil}
\affiliation{EPSRC National Epitaxy Facility, University of Sheffield, Sheffield S1 4DE, UK}

\author{I. Farrer}
\affiliation{Department of Electronic and Electrical Engineering, University of Sheffield, Sheffield S1 4DE, UK}

\author{J. Heffernan}
\affiliation{Department of Electronic and Electrical Engineering, University of Sheffield, Sheffield S1 4DE, UK}

\author{M.S.~Skolnick}
\affiliation{Department of Physics and Astronomy, University of Sheffield, Sheffield S3 7RH, UK}
\author{L.R. Wilson}
\affiliation{Department of Physics and Astronomy, University of Sheffield, Sheffield S3 7RH, UK}

\date{\today}

\begin{abstract}
Unidirectional photonic edge states arise at the interface between two topologically-distinct photonic crystals. Here, we demonstrate a micron-scale GaAs photonic ring resonator, created using a spin Hall-type topological photonic crystal waveguide. Embedded InGaAs quantum dots are used to probe the mode structure of the device. We map the spatial profile of the resonator modes, and demonstrate control of the mode confinement through tuning of the photonic crystal lattice parameters. The intrinsic chirality of the edge states makes them of interest for applications in integrated quantum photonics, and the resonator represents an important building block towards the development of such devices with embedded quantum emitters.
\end{abstract}

\pacs{}

\maketitle 
The integration of quantum photonic elements on chip presents a highly promising route to the realisation of scalable quantum devices. A key requirement of such an approach is the development of optical waveguides exhibiting low loss and negligible back scatter.\cite{Dietrich2016} Recently, topological waveguides have emerged as a new class of photonic device enabling the robust propagation of light on chip.\cite{Rechtsman2013,Barik2016,Barik2018,Yang2018a,Yang2018b,Wu2015,Khanikaev2017,Christiansen2019,Anderson2017} At the interface between two topologically-distinct photonic crystals (PhCs), counter-propagating edge states of opposing helicity arise, which are ideal for optical waveguiding.\cite{Lu2014,Ozawa2019,Yang2018a,Sun2019,Anderson2017} Significant developments in this field include the demonstration of efficient guiding of light around tight corners,\cite{Barik2018,Yamaguchi2019,Shalaev2019,He2019} robust transport despite the presence of defects,\cite{Hafezi2013} and integration with passive optical elements including nanobeam waveguides\cite{Shalaev2019} and grating couplers.\cite{Barik2018} Compatibility with embedded quantum emitters such as quantum dots (QDs) has been demonstrated, and used to probe the waveguide transmission.\cite{Barik2018,Yamaguchi2019} Recently, chiral coupling of a QD to a topological waveguide was demonstrated.\cite{Barik2018,barik2019chiral,mehrabad2019chiral} This is a result of the intrinsic helicity of the edge states and is of great interest for chiral quantum optics.\cite{Lodahl2017} 

\begin{figure}[h!]
\includegraphics[width=0.99\columnwidth]{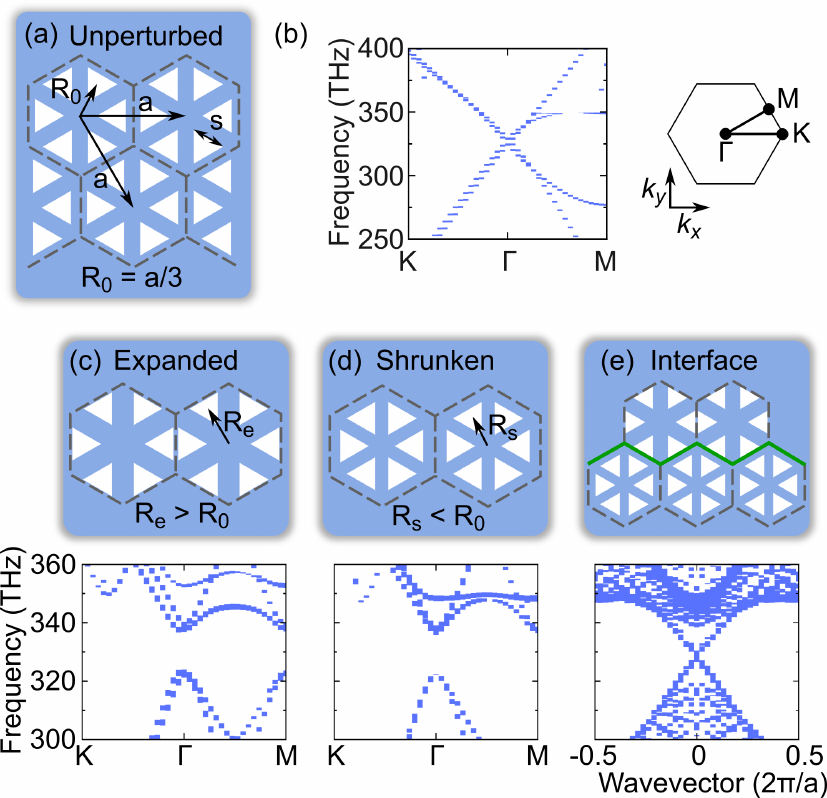}
\caption{\label{unit_cell}(a) Schematic showing the unperturbed photonic crystal (PhC). The triangles represent air holes. A two-dimensional PhC is formed using a triangular lattice of unit cells, with pitch $a$. (b) Band structure of the unperturbed lattice, revealing a Dirac cone at the $\Gamma$ point. Points of high symmetry in the Brillouin zone are shown in the accompanying schematic. (c,d) Schematic of (c) expanded and (d) shrunken unit cells. The band structure for a PhC formed from such a unit cell is also shown. In each case a bandgap is opened at the $\Gamma$ point. (e) Schematic of an interface formed between expanded and shrunken unit cells. Edge modes are seen to cross the bandgap in this case. The parameters used for the band structure calculations were: refractive index = 3.4, $h$=170nm, $s$=140nm, $a$=445nm, $R_e$=156nm and $R_s$=141nm.}
\end{figure}

Here, we use a spin Hall-type topological waveguide to create a GaAs topological photonic ring resonator, and probe its mode structure using embedded InGaAs QDs. We map the spatial dependence of the confined modes of the resonator, and demonstrate that perturbation of the PhC lattice can be used to tune the lateral confinement of the modes.

A schematic of the PhC forming the basis of our topological ring resonator is shown in Fig.~\ref{unit_cell}a. The unit cell of the PhC consists of six triangular air holes of side length $s$, etched into a GaAs membrane of thickness $h$. A two-dimensional PhC lattice is created using a hexagonal array of unit cells with period $a=s/0.31$. A key parameter of the lattice is the distance $R$ from the origin of the unit cell to the centre of each triangular aperture, with a graphene-like structure formed when $R=R_0=a/3$. We model the structure using a commercially available 3D finite-difference time-domain (FDTD) electromagnetic simulator,\cite{Lumerical} and show that in this case the PhC band structure exhibits a Dirac cone at the $\Gamma$ point, as shown in Fig.~\ref{unit_cell}b. 

However, when a perturbation is introduced such that $R\neq a/3$, a bandgap is opened at the $\Gamma$ point. This is shown for the case of PhCs formed using either expanded ($R_e>R_0$) or shrunken ($R_s<R_0$) unit cells in Fig.~\ref{unit_cell}c-d. Using FDTD, we determine the bandwidth of the PhC bandgap for a perturbation of either $R_e/R_0=1.05$ or $R_s/R_0=0.94$ to be $\sim$20THz ($\sim$55nm), centered at $\sim$330THz ($\sim$908nm). For the simulations, we took $h$=170nm and $s$=140nm. It is instructive to consider the nature of the bands in this case. For a shrunken (expanded) unit cell, the higher energy band has \textit{d}-(\textit{p}-) like character, and the lower energy band is \textit{p}-(\textit{d}-) like.\cite{Barik2016} This difference in character has an important consequence when an interface is realised between the two unit cells (see Fig.~\ref{unit_cell}e). The change in character of adjacent bands necessitates the formation of edge states, connecting bands of the same character across the interface. These can be clearly seen in the interface band structure shown in Fig.~\ref{unit_cell}e. The edge states exist within the bandgap, and the interface therefore supports confined modes which form the basis of a topological photonic waveguide.\cite{Parappurath2018}

\begin{figure}
\includegraphics[width=0.99\columnwidth]{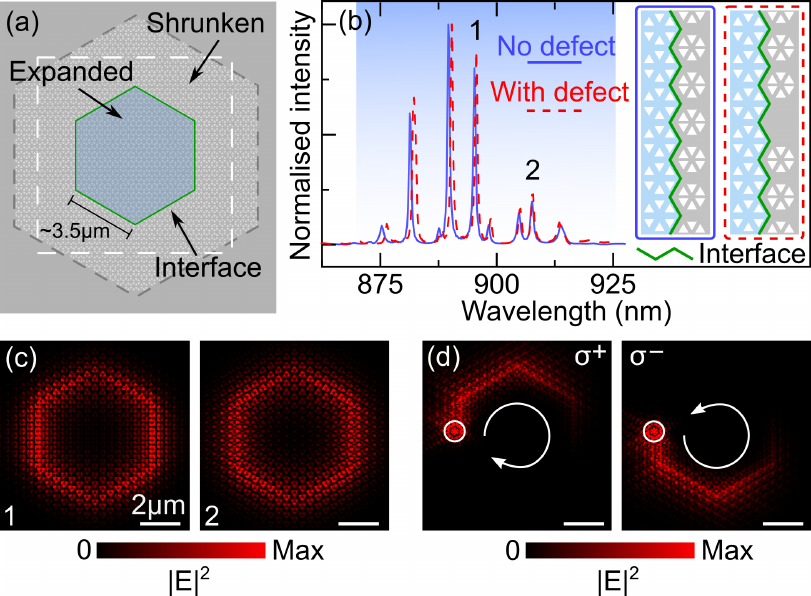}
\caption{\label{ring_resonator_sim}(a) Schematic of the ring resonator. The topological interface is indicated by the green line. (b) Mode spectrum at the resonator interface (blue solid line), determined using FDTD simulation. Resonator modes can be seen within the topological bandgap (shaded). The mode spectrum is also shown for a resonator containing a defect in the form of a single unit cell omitted from the interface (red dashed line). The inset shows a schematic of one side of the resonator, with and without the defect. (c) Spatially-resolved electric field intensity (linear scale) in the plane of the device, for the two modes numbered in (b) (without a defect). Confinement of light at the interface is clearly seen. The electric field intensity is evaluated within the white dashed region in (a). (d) Spatially-resolved electric field intensity (linear scale) in the plane of the device, for a (upper) $\sigma^+$ or (lower) $\sigma^-$polarised dipole source (position given by the open circle). The field intensity is averaged over the first 200fs of simulation time. Chiral emission is observed, with the direction of propagation (arrows) dependent on the dipole polarisation. The dipole is coupled to mode 1, as labelled in (b).}
\end{figure}

We harness the interface edge modes to create a photonic ring resonator. A hexagonal array of expanded unit cells ($R_e/R_0=1.05$) is embedded within a host array of shrunken unit cells ($R_s/R_0=0.94$), as shown schematically in Fig.~\ref{ring_resonator_sim}a. (In the following, this corresponds to an average perturbation of 5.5\%.) Internally, each side of the resonator is 8 unit cells in length, such that the total path length of the interface is $\sim21\mu m$. The hexagonal host array has an internal side length of 31 unit cells. FDTD simulations of the resonator reveal a characteristic spectral mode structure (Fig.~\ref{ring_resonator_sim}b), which lies within the topological bandgap (determined by monitoring the power radiated by a dipole source in an expanded-unit-cell PhC). The three most prominent modes have an average quality factor (Q factor) of 1760 (range 1600-1900). Spatial intensity profiles for two different modes (Fig.~\ref{ring_resonator_sim}c) confirm that they are confined to the topological interface.

The topological character of the resonator modes can be seen in protection against certain defects, and in the helical nature of the modes. We consider a defect in the form of a single unit cell missing from the resonator interface. This is seen to have little effect on the mode spectrum (dashed line in Fig.~\ref{ring_resonator_sim}b) and does not affect the mode Q factors. To demonstrate helicity of the modes, we position a circularly polarised dipole at the centre of a unit cell adjacent to the interface, and show that the propagation direction of the dipole emission is dependent on the handedness of the dipole polarisation (see Fig.~\ref{ring_resonator_sim}d). This highlights the potential of the resonator for chiral quantum optics \cite{Lodahl2017} and spin-photon interfaces.\cite{Coles2016,Javadi2018}

\begin{figure}
\includegraphics{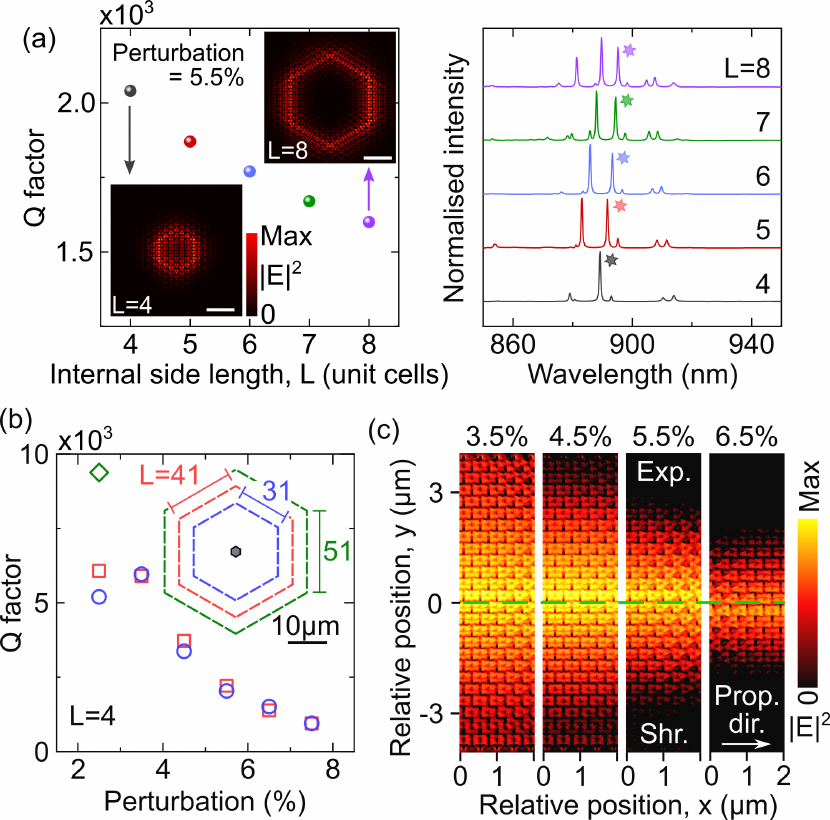}
\caption{\label{finesse} (a) Simulated resonator Q factor versus side length $L$ (the number of internal unit cells at the interface) for a PhC perturbation of 5.5\%. The size of the host PhC is fixed at 31 unit cells. Electric field intensity profiles (linear scale) for $L=4$ and $L=8$ are shown as insets (scale bars 2$\mu$m). Mode spectra for each case are shown to the right. The Q factor is evaluated for the mode marked with a star. (b) Simulated Q factor versus perturbation for a resonator with $L=4$. Blue circles (red squares, green diamond) correspond to a host PhC size of 31 (41, 51) unit cells. The inset shows the relative size of the resonator (black hexagon) and the host PhC in each case. (c) Simulated spatially-resolved electric field intensity for waveguides with perturbations of 3.5\% to 6.5\%, on a logarithmic intensity scale. The mode propagates in the \textit{x} direction (prop. dir.). The unit cell is expanded (Exp.) above and shrunken (Shr.) below the interface, which is highlighted by the dashed green line.}
\end{figure}

Next, we investigate the theoretical dependence of the resonator Q factor, first on the device dimensions, and then on the PhC perturbation. Fig.~\ref{finesse}a shows the mode spectra for five different resonators with internal side length between 4 and 8 unit cells. The host PhC has a side length of 31 unit cells and the perturbation is 5.5\%. The number of modes (mode spacing) increases (decreases) with increasing resonator size, as is expected for a Fabry-P\'erot-type resonator.\cite{Mark_Fox_2006} In each case, the Q factor of a mode near the centre of the distribution (marked with a star) is calculated. The Q factor is greater than 2,000 for the smallest resonator, and decreases slightly with increasing resonator size. Fixing the internal size of the ring at 4 unit cells, we then vary the perturbation, and see that the Q factor increases with decreasing perturbation, down to 3.5\% perturbation (blue circles in Fig.~\ref{finesse}b). The dependence of the Q factor on both the resonator size and perturbation can be understood as the result of the finite propagation length of the waveguide. This is due to scattering of the mode at the $\Gamma$ point, and subsequent coupling to free-space modes, the probability of which increases with increasing perturbation.\cite{Barik2016,Parappurath2018} For the smallest perturbation of 2.5\%, we find that the size of the host PhC must be increased to obtain the largest Q factor. Increasing the internal side length of the host initially to 41 unit cells (red squares in Fig.~\ref{finesse}b) has little effect for perturbations greater than 2.5\%, indicating that in this case the host PhC is already sufficiently large. However, for a host PC size of 51 unit cells (green diamond) a higher Q factor of $\sim$9,400 is obtained for a perturbation of 2.5\%. This is due to the mode confinement weakening as the perturbation decreases, such that the host PhC must be larger to prevent in-plane loss into the membrane surrounding the device (see Fig.~\ref{finesse}c).
 
\begin{figure}
\includegraphics[width=0.99\columnwidth]{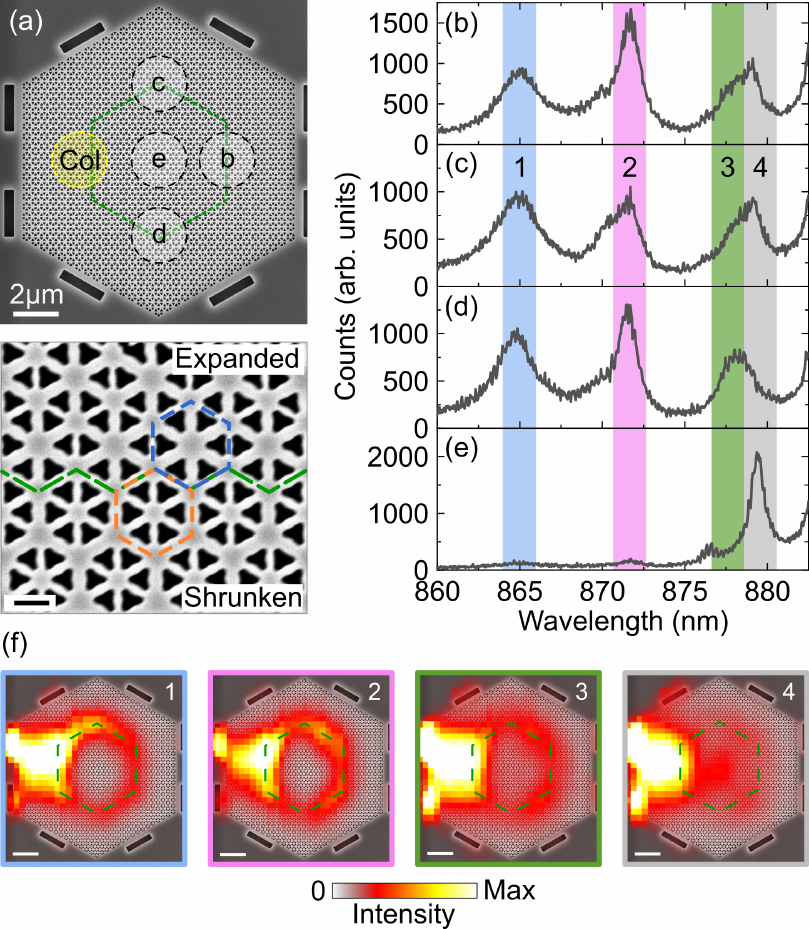}
\caption{\label{Results}(a) Scanning electron microscope (SEM) overview of the ring resonator (top) and close up of the waveguide interface (bottom, scale bar 200nm). (b-e) High power PL spectra for four different excitation locations on the device, as labelled in (a). In each case, the collection spot was fixed on the left hand side of the resonator (yellow circle labelled `Col' in (a)). (f) Spatially-resolved, integrated PL intensity maps for four different bandwidths, as numbered and colour-coded in (b-e). The zero of the linear colour scale is transparent. An SEM image of the device is positioned under each map, and the interface highlighted with a dashed green line. Scale bars 2$\mu$m.}
\end{figure}

Experimentally, we fabricate topological ring resonators in a nominally 170nm-thick GaAs membrane, using standard electron beam lithography and dry etching techniques. A scanning electron microscope (SEM) image of a representative ring resonator is shown in Fig.~\ref{Results}a. The resonator has an internal side length of 8 unit cells, and is embedded in a host PhC with a side length restricted to 16 unit cells due to experimental limitations. The simulated Q factor in this case is reduced to a maximum of $\sim$870 for a perturbation of 5.5\%, due to additional loss into the membrane. Devices are fabricated with perturbation between 2.5\% ($R_e/R_0=1.02$, $R_s/R_0=0.97$) and 5.5\% ($R_e/R_0=1.05$, $R_s/R_0=0.94$). 

To characterise the devices, the sample is mounted in an exchange gas cryostat, and the waveguide modes of the resonator are probed using micro-photoluminescence ($\mu$-PL) spectroscopy. We use an excitation wavelength of 808nm, and focus the laser to a spot size of $\sim2\mu$m. High power excitation is used to generate broadband emission from the QD ensemble. Mirrors in the collection path, with motorized adjusters, enable the collection of PL emission from a location either coincident with, or spatially distinct from, the excitation spot. We first position the excitation laser spot on one side of the ring (region `b' in Fig.~\ref{Results}a), and detect light emitted from a location on the opposite side of the resonator (region `Col' in Fig.~\ref{Results}a). This enables us to detect light coupled to the interface, whilst rejecting PL which is emitted into free-space modes. The PL spectrum for a ring resonator with a unit cell perturbation of 5.5\% is shown in Fig.~\ref{Results}b. Resonator modes with a period of $\sim7$nm are observed, consistent with simulation. The modes have Q factors in the range 200-500, possibly limited by dopant-related absorption in the GaAs membrane. The wavelength range in which the modes are observed is slightly shorter than in simulation, an effect we attribute to larger fabricated hole sizes than designed. The short wavelength edge of the bandgap therefore lies at a wavelength outside the bandwidth of the QD emission, such that the bandwidth of the topological bandgap cannot be determined precisely.

We next excite PL at several other positions on the resonator interface while collecting from the same location as previously, and observe that the mode structure remains unchanged (see Fig.~\ref{Results}c-d). However, when the excitation location is distinct from the interface (for instance, at the very centre of the resonator), the PL spectrum is quite different, and the modes observed previously at the interface are absent (Fig.~\ref{Results}e)). This is strong evidence that the modes are confined to the interface. (A different peak is seen in Fig.~\ref{Results}e. We show below that this is distinct from the resonator modes.)

To visualise the modes more clearly, we raster scan the excitation laser across the device, simultaneously acquiring PL spectra from the same (fixed) collection spot used above. For each excitation position on the device, we sum the measured PL intensity over three different bandwidths corresponding to the different resonator modes seen in Fig.~\ref{Results}b-d. The resulting PL intensity maps are shown in Fig.~\ref{Results}f. Several features are apparent in the data. Most significantly, PL emission is observed from all positions along the resonator interface, with a clear node in the centre of the resonator, showing that light is guided along the interface. A secondary feature is the bright region on the left of each intensity map, which corresponds to the PL collection location. This is the result of a fraction of the QD PL emission coupling to free-space modes. Such emission is collected only when the excitation laser is positioned close to the collection spot. The PL maps also serve to highlight the location of trenches used to aid fabrication of the device (for instance, the two bright regions to the left of the collection location). These were used to accurately position the device schematic over each PL map. In Fig.~\ref{Results}f we also show an intensity map integrated over the single peak seen in the PL spectrum from the centre of the device. The map is quite different to those for the resonator modes seen at the interface, as this peak is not related to guided modes at the interface.

\begin{figure}
\includegraphics[width=0.99\columnwidth]{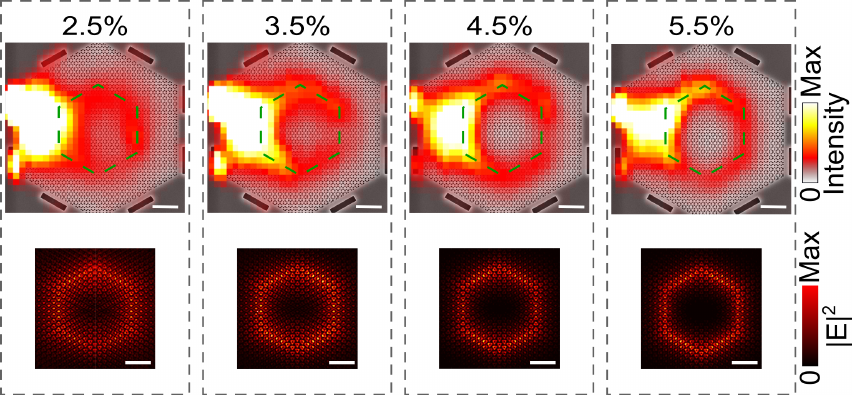}
\caption{\label{Perturbation dependence}Dependence of waveguide mode transverse confinement on perturbation. Upper panels: Spatially-resolved, integrated PL intensity maps for different unit cell perturbations (as labelled). The integration is for a mode centered at $\sim865$nm in each case. The zero of the linear colour scale is transparent. An SEM image of the device is positioned under each map, and the interface highlighted with a dashed green line. Lower panels: Simulated spatially-resolved electric field intensity (linear scale) of a representative ring resonator mode, for devices with the same perturbations as in experiment. Scale bars 2$\mu$m.}
\end{figure}

Finally, we investigate the effect of the unit cell perturbation on the spatial confinement of the resonator modes. Fig.~\ref{Perturbation dependence}a shows an integrated PL intensity map for each of four different ring resonators, with unit cell perturbation increasing from 2.5\% to 5.5\%. In each case, the intensity map corresponds to a single resonator mode, centered at $\sim 865$nm. A clear trend is observed, with the mode confinement being strongest for the largest perturbation, and weakening with decreasing perturbation. For a perturbation of 5.5\%, the mode decays over a distance of 2.7$\pm$0.1$\mu$m, transverse to the propagation direction. This value increases to 4.1$\pm$0.2$\mu$m for a perturbation of only 2.5\%. The change in confinement is consistent with FDTD simulations (see Fig.~\ref{finesse} and Fig.~\ref{Perturbation dependence}b), which show that, as the perturbation is decreased, the waveguide mode increasingly extends into the bulk PhC. From the simulations, we estimate that the spatial extent of the mode normal to the propagation direction increases from $\sim820$nm for a perturbation of 5.5\% to $\sim1640$nm for a perturbation of 2.5\%. (The discrepancy between experiment and simulation is due to convolution of the experimental data with the laser spot size of $\sim2\mu$m.) This suggests a robust method to tune the degree of evanescent coupling between the ring resonator and an adjoining bus waveguide, for instance in an add-drop filter. The perturbation is dependent on the location of the triangular apertures forming the PhC. This is simple to control lithographically, unlike the case of devices which rely on fine tuning of the resonator-waveguide spacing.

In conclusion, we have created a GaAs spin Hall topological photonic ring resonator, and used embedded InGaAs QDs to probe the mode structure of the device. Using spatially resolved PL measurements, we demonstrated that the modes were confined to the PhC interface. Furthermore, we showed that by controlling the perturbation of the PhC unit cell, the spatial confinement of the modes could be tuned. The resonator represents an important building block in the development of integrated photonic devices using embedded quantum emitters.

Data supporting this study are openly available from the University of Sheffield repository. \footnote{\MakeLowercase{h}ttps://doi.org/10.15131/shef.data.9944957}

This work was supported by EPSRC Grant No. EP/N031776/1. 

\bibliography{Main}

\end{document}